\documentclass[aps,prl,reprint,groupedaddress]{revtex4-1}

\usepackage{graphicx}
\usepackage{dcolumn}
\usepackage{bm}
\usepackage{MnSymbol}	
\usepackage{wasysym}%
\usepackage{amsmath} 

\newcommand{\be}{\begin{equation}}
\newcommand{\ee}{\end{equation}}
\newcommand{\bee}{\begin{eqnarray}}
\newcommand{\eee}{\end{eqnarray}}

\begin{document}

\preprint{APS/123-QED}

\title{Vortices enhance diffusion in dense granular flows}

\author{Prashidha Kharel}
\email{prashidha.kharel@sydney.edu.au}
\affiliation{Particles and Grains Laboratory, School of Civil Engineering, The University
of Sydney, Sydney, NSW 2006, Australia.}
\author{Pierre Rognon}
\email{pierre.rognon@sydney.edu.au}
\affiliation{Particles and Grains Laboratory, School of Civil Engineering, The University
of Sydney, Sydney, NSW 2006, Australia.}

\date{\today}
             
\begin{abstract}
This Letter introduces unexpected diffusion properties in dense granular flows, and shows that they result from the development of partially jammed clusters of grains, or granular vortices. Transverse diffusion coefficients $D$ and average vortex sizes $\ell$ are systematically measured in simulated plane shear flows at differing inertial numbers $I$ revealing (i) a strong deviation from the expected scaling $D\propto d^2 \dot \gamma$ involving the grain size $d$ and shear rate $\dot \gamma$ and (ii) an increase in average vortex size $\ell$ at low $I$, following $\ell\propto dI^{-\frac{1}{2}}$ but limited by the system size. A general scaling $D\propto \ell d \dot \gamma $ is introduced that captures all the measurements and highlights the key role of vortex size. This leads to establishing a scaling for the diffusivity in dense granular flow as $D\propto d^2 \sqrt{\dot \gamma/ t_i}$ involving the geometric average of shear time $1/\dot\gamma$ and inertial time $t_i$ as the relevant time scale. 
Analysis of grain trajectories further evidence that this diffusion process arises from a vortex-driven random walk.
\end{abstract}

\pacs{ }

\maketitle

\setlength{\belowdisplayskip}{6pt} \setlength{\belowdisplayshortskip}{6pt}
\setlength{\abovedisplayskip}{0pt} \setlength{\abovedisplayshortskip}{0pt}

Shear-induced diffusion underlies key properties related to mixing, heat transfer and rheology of granular materials \cite{hsiau93,utter04}. For instance, it counterbalances size segregation in polydispersed granular flow down a slope, thereby affecting the dynamics of landslides and snow avalanches \cite{rognon07,marks12,schlick16}. It also produces strong convective heat fluxes across granular sheared layers \cite{rognon2010thermal}.

In dense granular flows, shear-induced diffusion is usually modelled by expressing a diffusivity $D \text{ [m}^2\text{/s]}$ in terms of the natural length and time scales, the grain size $d \text{ [m]}$ and shear rate $\dot \gamma\text{ [s}^{-1}]$ \cite{hsiau93,utter04,wandersman12,Fan15}:

\be \label{eq:simple}
D \propto d^2 \dot \gamma
\ee
The typical grain trajectory leading to this scaling is a random walk with a step of the order of $d$ and random changes in direction at a typical frequency $\dot \gamma$. However, the internal kinematics of dense granular flows do not simply follow such a random walk.

Firstly, the typical duration of grain rearrangement -or plastic events- within the flows was shown to correspond to a time scale $t_i = d \sqrt{\frac{\rho}{P}}$, named \text{inertial time}, involving the grain density $\rho$ and the normal stress $P$. Dense granular flows were found to correspond to small values of the inertial number $I \ll 1$, defined as \cite{da2005rheophysics, midi2004dense} :

\be
I = \dot \gamma t_i.
\ee

\noindent Interestingly, small values of $I$ implies that the time scale for these grain rearrangements is much shorter than the shear time, $t_i \ll \dot \gamma ^{-1}$. 
Secondly, grain kinematics in dense granular flows may exhibit strong spatial correlations for short periods of time \cite{radjai_turbulentlike_2002, pouliquen2004velocity, miller2013eddy, woldhuis2015fluctuations,rognon2015long,kozicki2017investigations,saitoh2016anomalous,*saitoh2016enstrophy}. 
Whether and how these kinematic features could affect the diffusive property, in particular the scaling (\ref{eq:simple}) remains poorly explored.

The purpose of this Letter is to establish the scaling of the diffusivity $D$ in dense granular flows, and to identify its origin in terms of internal kinematics. \\

\begin{figure}[!b]
    \centering
	\includegraphics[width=0.9\linewidth]{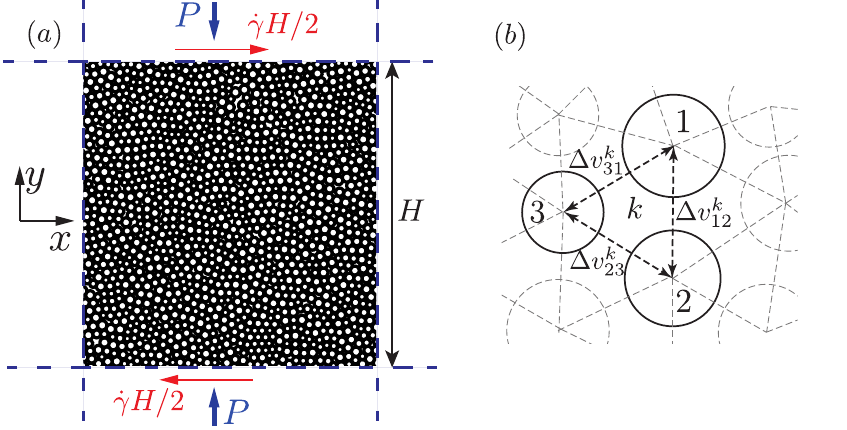}
    \centering
    \caption{(a) Plane shear geometry with biperiodic boundary conditions (dashed lines). (b) Schematic of Delaunay triangulation and computation of relative velocities for triplet of grains.} 
    \label{fig1}
\end{figure}

\textit{Measuring shear-induced diffusion.}\textemdash  
We used a discrete element method to simulate plane shear flows in a bi-periodic domain prescribing both the shear rate $\dot \gamma$ and normal stress $P$ (see Figure \ref{fig1}(a)). This flow geometry avoids any stress gradient and walls, and thus produces homogeneous flows in which shear rate, shear and normal stresses, and inertial number $I$ are spatially constant. This enables tracking grains diffusing across the shear direction over long distances, while they always experience the same flow conditions. 

Simulated grains are disks which diameters are uniformly distributed in the range $d\pm 20\%$ in order to avoid crystallisation. They interact \textit{via} dissipative, elastic and frictional contacts characterised by a Young's modulus $E = 1000 P$, a normal coefficient of restitution $e=0.5$, and a friction coefficient of $0.5$ for frictional grains and $0$ for frictionless grains. In the dense regime, the value of $E$ and $e$ do not significantly affect the flow properties \cite{da2005rheophysics}. 

This method was used to simulate flows of differing inertial numbers within the range $7\times10^{-4}\leq I \leq 5\times10^{-1}$, in systems of differing size in the range $20 \leq H/d \leq 120$. The numerical experiments start by randomly placing grains with no contact. Then, normal stress $P$ and shear rate $\dot\gamma$ are simultaneously applied until a steady state is reached. All measurements reported below are performed over a period of time of $25/\dot\gamma$ corresponding to $25$ shear deformations, after the flow has reached a steady state.

\begin{figure}[!t]
    \centering
	\includegraphics[width=\linewidth]{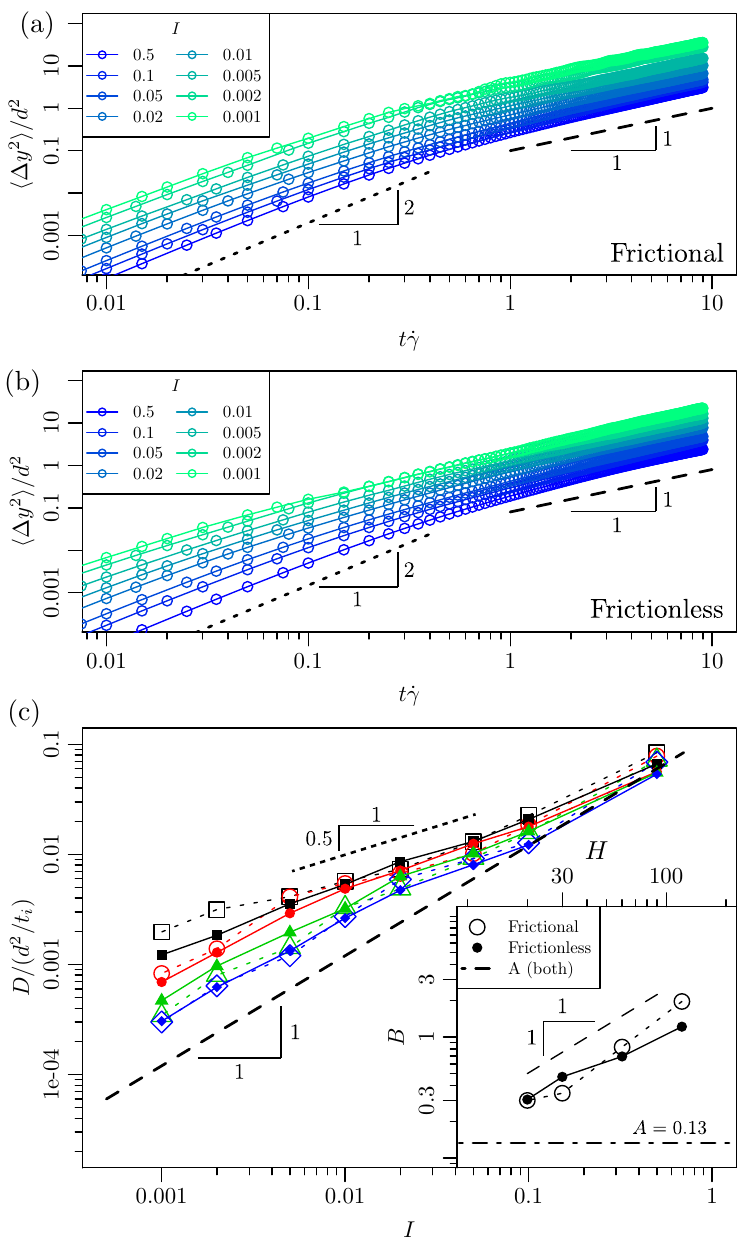}
    \centering
    \caption{Time evolution of mean square displacement $\langle\Delta y^2\rangle$ for different values of inertial number $I$ for (a) frictional grains and (b) frictionless grains   (system size $H/d = 120$). (c) Measured diffusivity $D$ versus inertial number $I$ for systems of differing size with $H/d$ = 20 (diamonds), 30 (triangles), 60 (circles), 120 (squares). Hollow and filled symbols represent frictional and frictionless grains, respectively. Inset: Variation of the constant $B$ (see text) with different system size.} 
    \label{fig2}
\end{figure}

The primary quantity of interest in quantifying the shear-induced diffusion is the mean square displacement of grains across the shear direction, defined as: $\langle \Delta y^2 \rangle (t) = \frac{1}{N} \sum_{i=1}^N (y_i(t=0) -y_i(t))^2$, where the sum runs on all the grains in the shear cell.  A \textit{normal}-diffusive behaviour, such as induced by a 1d-random walk, would lead to a mean square displacement increasing linearly with time, 
$\langle \Delta y^2 \rangle (t) = 2 Dt$. By contrast, grains moving up or down at a constant speed $v$ would lead to a quadratic scaling $\langle \Delta y^2 \rangle (t) = v t^2$, corresponding to a \textit{super}-diffusive behaviour.

Figures \ref{fig2}(a-b) show the time evolution of mean square displacements for  frictional and frictionless grains, and for different values of the inertial number $I$. They evidence the existence of a super-diffusive behaviour until flows undergo a fraction of shear deformations, followed by a normal diffusive behaviour at larger deformation - a result consistent with previous observation \cite{radjai_turbulentlike_2002,utter04,wandersman12,Fan15}. Figure \ref{fig2}(c) shows the value of diffusivity $D$ normalised by $d^2/t_i$, obtained by fitting the mean square displacement at larger deformations ($t\dot \gamma>2$) by the function $2 Dt$. Thus normalised, the scaling (\ref{eq:simple}) would predict $\frac{D}{d^2/t_i} \propto I$.

\noindent By contrast, figure \ref{fig2}(c) reveals a significant breakdown of the scaling (\ref{eq:simple}) at low inertial numbers. Instead, three regimes appear.  At large $I$, the expected scaling (\ref{eq:simple}) is recovered, with $D \approx A d^2 \dot \gamma$ and $A\approx 0.13$. For the lowest values of inertial number, the scaling (\ref{eq:simple}) is also recovered, with $D \approx B d^2 \dot \gamma$, however, it then involves a constant $B$ much greater than $A$, which seemingly increases linearly with the system size (see inset of figure \ref{fig2}(c)). For intermediate inertial numbers, the scaling (\ref{eq:simple}) is not valid. The diffusivity is then no longer directly proportional to the shear rate, for both frictional and frictionless grains.


These results show that the diffusivity at low inertial number are significantly larger than the prediction of (\ref{eq:simple}) measured for high $I$. They further indicate that this enhancement becomes more acute for systems with larger width $H$.


\begin{figure*}[!t]
    \centering
	\includegraphics[width=0.78\textwidth]{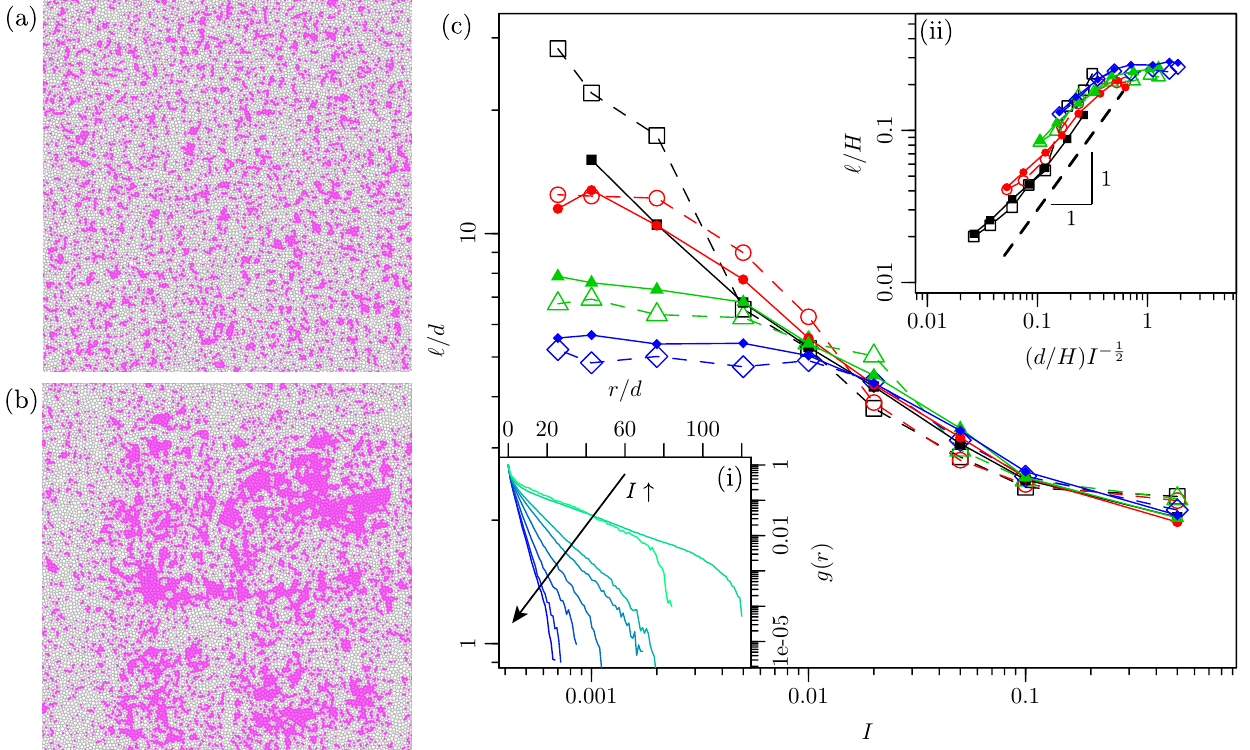}
    \centering
    \caption{Granular vortices. (a,b) Snapshots illustrating granular vortices, i.e. triplets of kinematically jammed grains shown in magenta for $I=$ 0.1 (a) and 0.001 (b) for $H/d=120$ system with frictional grains. (Movies available online \cite{Note1})
    (c) Average vortex size $\ell$ \text{versus} inertial number $I$ for frictional grains and frictionless grains in systems of differing size $H$. Symbols represent same configurations as those in figure \ref{fig2}(c). Inset (i): Correlation function $g(r)$ measured in flows of differing inertial numbers ranging from 0.5 to 0.001 for $H/d=120$ system with frictional grains. Inset (ii): Same plot as (c) but rescaled with the system size $H$ and a power of inertial number. } 
    \label{fig3}
\end{figure*}

\textit{Granular vortices length scale $\ell$}\textemdash
Let us now seek to rationalise the measured diffusivities by analysing the grain kinematics within the flows. 
One approach would be to express diffusivity in terms of velocity fluctuations $\delta v$, which leads to a scaling of the form $D\propto d \delta v$. 
However,  in the supplemental material 
\footnote{See supplemental material at [URL for supplemental paper] for more details on diffusivity in terms of velocity fluctuations, discussions on alternative approaches on measuring vortex size, and movies showing time evolution of vortices and mixing of differently coloured layers of grains for two inertial numbers}, 
we show that this scaling cannot fully capture the diffusivity, with its prediction getting worse for smaller systems and frictionaless grains.
Instead, we introduce in the following an analysis based on the existence and size of jammed clusters of grains, which are hereafter referred as \textit{granular vortices}.


In this aim, we developed the following method to detect vortices, involving two steps: (i) identifying at any point in time triplets of neighbouring grains considered kinematically jammed, and (ii) aggregating jammed triplets sharing two common grains into a larger vortex. Triplets of neighbouring grains are identified using a Delaunay triangulation. The criteria used to define whether a triplet is jammed is based on the relative normal velocity of the three pair of grains  $|\Delta v^k|_{max} = max(|\Delta v_{12}^k|;|\Delta v_{1 3}^k|;|\Delta v_{23}^k|)$ (see Fig. \ref{fig1}(b)). A triplet is considered kinematically jammed if all of the three relative normal velocities are smaller than the average relative normal velocity $\langle |\Delta v| \rangle$ between all pair of neighbouring grains in the full system, i.e. triplet $k$ is jammed if $|\Delta v^k|_{max} < \langle |\Delta v| \rangle$. The outputs of this method include a list of vortices at any point in time, each being comprised of a list of jammed grains. Snapshots of jammed grains are shown in figure \ref{fig3}(a,b). Alternative vortex detection methods and more details on the method used here are discussed in the Supplemental Material at \cite{Note1}.

In order to quantify the average vortex size, we define a correlation function $g(r)$ measuring the probability that two grains at a distance $r$ belong to the same vortex. $g(r)$ is calculated using randomly picked one hundred snapshots.
Figures \ref{fig3}(c-i) shows that, for smaller inertial numbers, these correlation functions exhibit a higher tail, indicating a higher probability of finding large clusters. The average vortex size $\ell$ is then deduced from $g(r)$ as:

\be \label{eq:avg_cluster}
\ell = \frac{\sum_r (r+d)  g(r)}{\sum_r  g(r)}.
\ee

\noindent The average on $r+d$, rather than $r$ is chosen to represent the distance between the centre of the two grains $r$ plus their outer halves $d$.

Figure \ref{fig3}(c) shows the average vortex size measured in flows of differing inertial number $I$ and widths $H$. At high inertial numbers ($I \gtrsim 0.1$), it appears that the vortex size is nearly constant and close to one grain size ($\ell\approx d$). At intermediate inertial numbers, it seemingly scales like:

\be \label{eq:l}
\frac{ \ell}{d} \propto I^{-\frac{1}{2}}
 \ee
 
\noindent  It ultimately reaches a maximum value $\ell^{max}$ at smaller inertial numbers, which is proportional to the system size, $l^{max} \approx H/4$.
The scaling (\ref{eq:l}) and system size effects are further highlighted on figure \ref{fig3}(c-ii). These observations indicate that while vortices tend to grow as the inertial number decreases, this growth is eventually limited by the system size. 

\begin{figure}
    \centering
	\includegraphics[width=\linewidth]{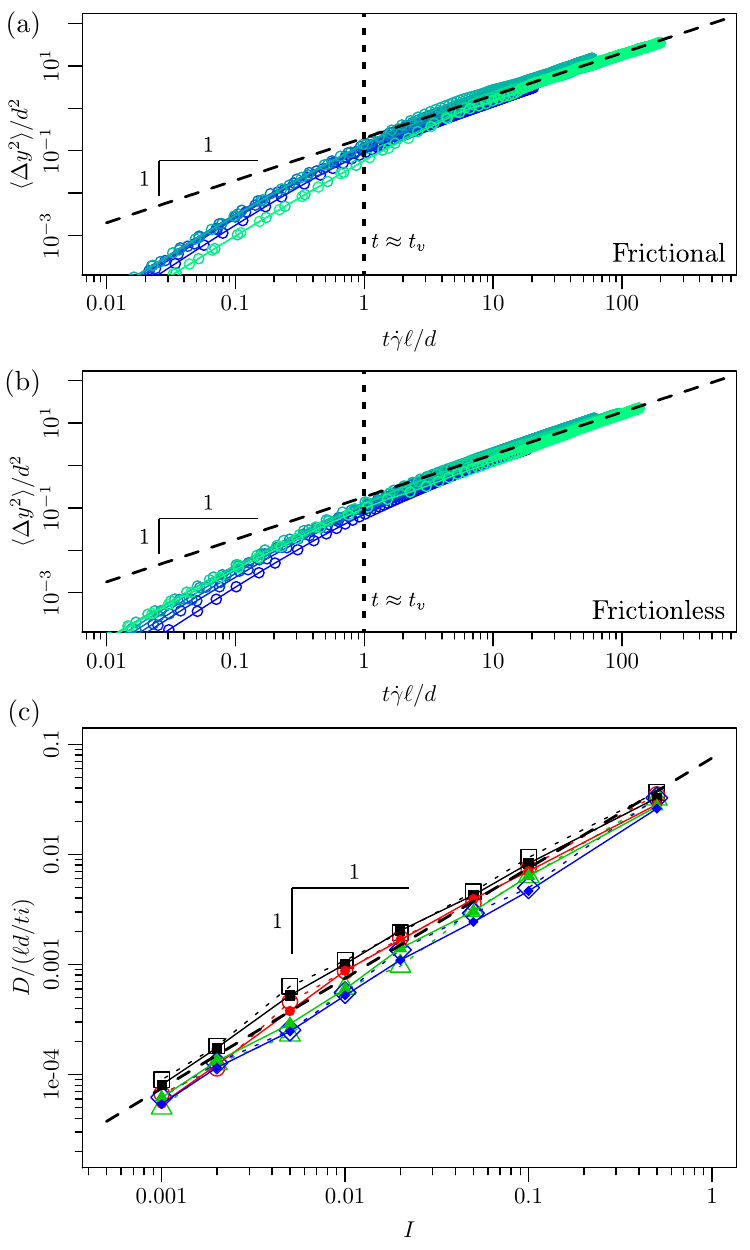}
    \centering
    \caption{Time evolution of mean square displacement $\langle\Delta y^2\rangle$ for different values of inertial number $I$ for frictional grains (a) and frictionless grains (b) of system size $H/d = 120$. (c) Measured diffusivity $D$ versus inertial number $I$ and vortex size $\ell$ for systems of differing size $H$. Dashed line represents the function $D/(d \ell /t_i) = 0.08 I$. All the symbols represent the same configurations as those in figure \ref{fig2}.} 
    \label{fig4}
\end{figure}

\textit{Scaling of diffusivity with $\ell$}\textemdash
Figures \ref{fig4} (a,b) shows a collapse of the mean square displacements in the normal-diffusive regime when plotted as a function of $t \dot\gamma \ell/d$ instead of $t \dot\gamma$. Consistently, figure \ref{fig4}(c) shows a linear increase of the ratio $\frac{D}{d \ell /t_i}$ with the inertial number $I$, which leads to the following scaling for the diffusivity:

\be \label{eq:new_diff}
D \propto d \ell \dot\gamma.
\ee

\noindent This scaling captures all the measured diffusivities in flows of differing inertial numbers $I$ and width $H$. 

Introducing the scaling (\ref{eq:l}) of the vortex size into (\ref{eq:new_diff}) leads to the  following expression for the diffusivity:

\be \label{eq:new_diff_I}
D \propto d^2 \sqrt{\frac{\dot \gamma}{t_i} },
\ee

\noindent which applies when the vortex size is not limited by the system size and
involves the geometric average of the shear time $1/\dot\gamma$ and inertial time $t_i$.
When the vortex size is limited by the system size, i.e. when $I \lesssim \left(\frac{d}{H}\right)^2$ according to (\ref{eq:l}), the diffusivity would scale like $D \propto d\ell^{max} \dot \gamma \approx d H \dot \gamma$, consistent with figure \ref{fig2}(c-inset).

\textit{A vortex-induced random walk} \textemdash
A random walk of step $s$ and frequency $f$ would lead to a mean square displacement exhibiting a super diffusive behaviour ($\langle \Delta y^2 \rangle (t) = (sf)^2 t^2$) at times shorter than $f^{-1}$, reaching the value $\langle \Delta y^2 \rangle (f^{-1}) = s^2$ at time $f^{-1}$, and following a normal diffusive behaviour afterward ($\langle \Delta y^2 \rangle (t) \propto 2 s^2f t$). This is precisely the pattern evidenced on figure \ref{fig4}(a,b), which shows a transition from super-diffusive to normal-diffusive behaviour at time $t \approx \frac{d}{\ell \dot \gamma}$, with a mean square displacement reaching a value of approximately $(0.3d)^2$. This suggests that grain trajectories in dense granular flows follow a random walk of typical step $s\approx0.3d$ and frecquency $\frac{\ell}{d} \dot \gamma$, which may be orders of magnitude higher than $\dot \gamma$.

This observation is consistent with a scenario involving vortices of size $\ell$ rotating at speed $\dot \gamma$ for a short amount of time \cite{griffani2013rotational}. Grains at the vortex periphery would then move at an average speed $\ell \dot\gamma$
and take a time $d/(\ell \dot\gamma)$ to cover a distance of the order of $d$, leading to the super-diffusive regime.
After this period of time, these grains would become part of a different vortex which would make them move a different direction, producing a new step for the random walk. This scenario points out a typical vortex life-time $t_v$ scaling like $d/(\ell \dot\gamma)$. 
In the dense regime, when vortices are not affected by the system size, this vortex life-time can be expressed as:

\be \label{eq:tv}
t_v \propto \frac{d}{\dot\gamma\ell} \propto \sqrt{\frac{t_i}{\dot \gamma} }.
\ee

\noindent The diffusivity (\ref{eq:new_diff_I}) can then be expressed as a function of the vortex life time as:

\be \label{eq:new_diff_t_v}
D \propto \frac{d^2}{t_v}.
\ee

\noindent Accordingly, one can interpret shear-induced diffusivity in dense granular flow to arise from a random walk with a step of the order of $d$, and with a frequency $t_v^{-1}$ controlled by the vortex life-time rather than by the shear rate $\dot \gamma$. As a result, during one shear deformation time ($1/\dot\gamma$) there will be $1/(t_v\dot\gamma) = \ell/d$ number of vortex random steps, which ultimately results in the enhancement of the diffusion. This contrasts with eq (\ref{eq:simple}) which assumes a single random step per shear deformation.

\textit{Conclusion}\textemdash
This study shows that the existence of vortices greatly affects shear-induced diffusion in dense granular flows. 
A key scaling law  ($\ref{eq:new_diff}$) is evidenced that relates the diffusion $D$ to the vortex size. 
This scaling law could help rationalising shear-induced diffusive behaviours in other glassy materials such as dense suspensions \cite{during2014length,*peters2016direct}, Lennard-Jones glasses \cite{lemaitre2009rate,*PhysRevE.60.3107,*heussinger2010superdiffusive}, foams \cite{durian_foam_1995,*debregeas2001deformation} and emulsions \cite{goyon_spatial_2008,*nordstrom2011dynamical}, which also exhibit correlated kinematics fields. 

For unbounded dense granular flows, a second scaling  (\ref{eq:l}) is evidenced that relates the vortex size to the inertial number. These two scalings indicate that the diffusivity is then directly defined by the grain size and a geometric average of the shear and inertial time, as in (\ref{eq:new_diff_I}). This expression can readily be used in continuum hydrodynamic models of dense granular flows.

Finally, the analysis of the grain trajectory indicates that granular vortices are short lived, with a typical life time given by (\ref{eq:tv}) that can be orders of magnitude shorter than the shear time. This time scale is reminiscent of velocity auto-correlation decay time \cite{PhysRevE.81.040301,*PhysRevE.91.062206,*PhysRevE.94.012904} and various rearrangement time scales \cite{PhysRevE.58.3515,*PhysRevLett.89.095704,*Keys:2007aa} measured in other glassy systems. Besides explaining diffusive properties, this feature could help rationalise rheological properties of dense granular flows such as non-local behaviours \cite{komatsu2001creep,*kamrin2012nonlocal,*bouzid2013nonlocal,rognon2015long}.
\nocite{da2005rheophysics,lerner2012unified,peyneau2008frictionless,PhysRevE.94.012904,rognon2015circulation,pouliquen2004velocity,saitoh2016anomalous,saitoh2016enstrophy}


%

\onecolumngrid
\break
\null
\thispagestyle{empty}
\includegraphics[page=1,width=\textwidth]{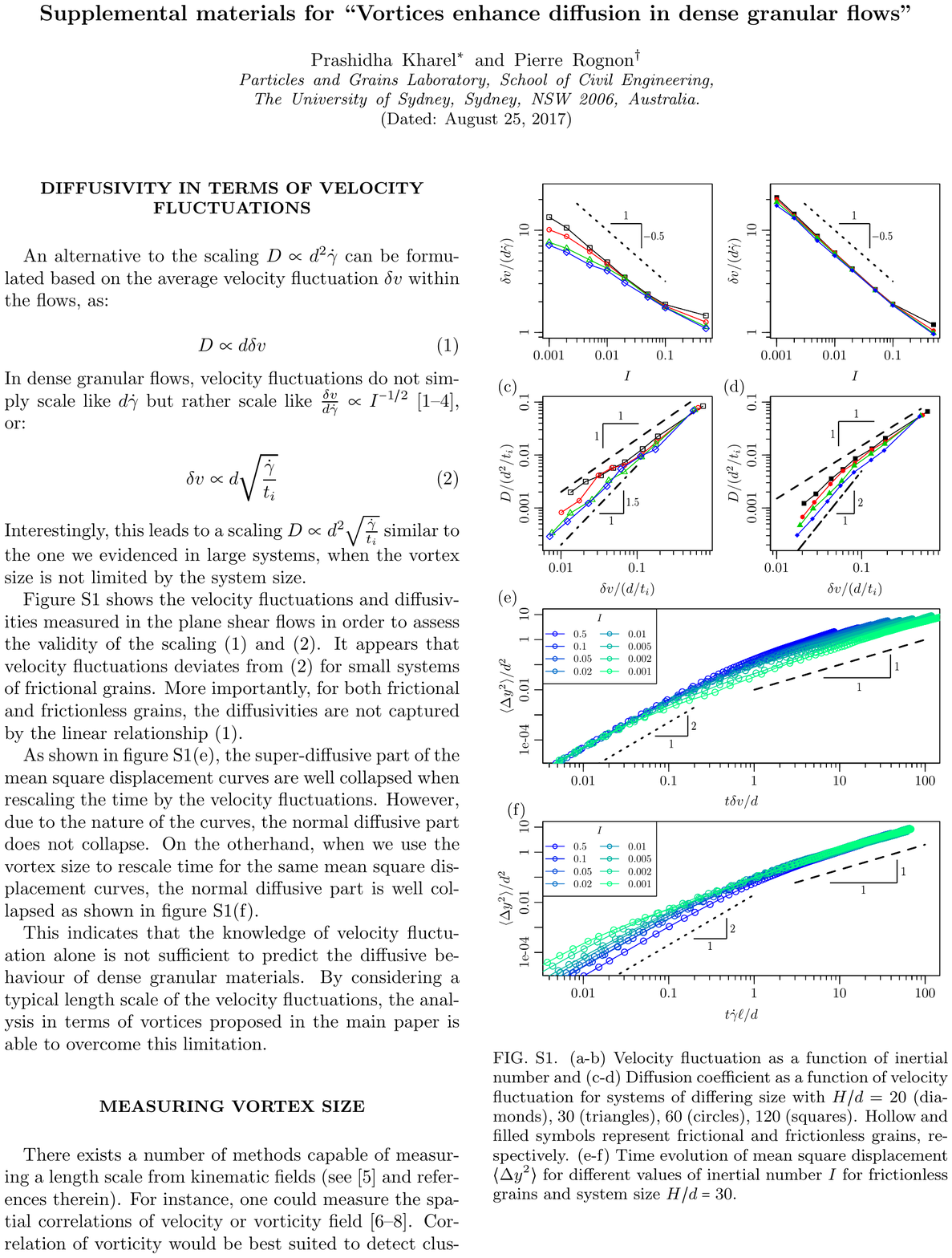}
\onecolumngrid
\break
\thispagestyle{empty}
\includegraphics[page=2,width=\textwidth]{diffusion_sup.pdf}

\end{document}